\documentclass[aps,prc,reprint,onecolumn,preprintnumbers,superscriptaddress,nofootinbib,notitlepage]{revtex4-1}
\usepackage{amsmath,amssymb}
\usepackage[latin1]{inputenc}
\usepackage[]{caption2}
\usepackage{graphics}
\usepackage[demo]{graphicx}
\usepackage{hyperref}
\usepackage{lipsum}
\usepackage{wrapfig, blindtext}
\usepackage{comment}
\usepackage{tabularx}
\usepackage{float}

\newcommand{\be}{\begin{eqnarray}}
\newcommand{\ee}{\end{eqnarray}}

\newcommand{\ml}{\mathcal}
\newcommand{\bs}{\boldsymbol}
\newcommand*\diff{\mathop{}\!\mathrm{d}}

\newcommand{\nn}{\nonumber}

\begin{document}

\title{Quarkonium Suppression in the Open Quantum System Approach}

\author{Xiaojun Yao}
\email{xjyao@mit.edu}
\affiliation{Center for Theoretical Physics, Massachusetts Institute of Technology, Cambridge, MA 02139 USA}

\begin{abstract}
Quarkonium suppression in relativistic heavy ion collisions has been studied experimentally for decades to probe the properties of the quark-gluon plasma. For this purpose, complete theoretical understanding of the time evolution of quarkonium inside the quark-gluon plasma is needed but challenging. 
Here I review recent progress in applying the open quantum system framework to describe the real-time dynamics of quarkonium, with a focus on the gauge-invariant chromoelectric field correlators of the plasma that control the dynamics.
\end{abstract}

\preprint{MIT-CTP/5393}
\maketitle

\section{Introduction}
Quarkonia are hadronic bound states containing a heavy quark-antiquark pair ($Q\bar{Q}$). The mass spectrum of ground and lower excited quarkonium states can be well described by nonrelativistic Schr\"odinger equations with phenomenological potential models such as the Cornell potential. When quarkonium states are placed inside the hot quark-gluon plasma (QGP), they can be broken into unbound heavy quark pairs due to plasma screening effects~\cite{Matsui:1986dk}. As a result, quarkonium production in relativistic heavy ion collisions will be suppressed if a QGP is created in the collisions, compared to that in proton-proton collisions. This motivates the experimental measurements of quarkonium production in heavy ion collisions and using quarkonium suppression as a diagnostic signature and a probe of the QGP. Recent reviews can be found in Refs.~\cite{Rothkopf:2019ipj,Chapon:2020heu,Muller:2021ygo}

To study properties of the QGP by using quarkonium, theoretical inputs are necessary in addition to the experimental measurements. The real-time dynamics of quarkonium inside the QGP is complicated since the description has to account for plasma screening, dissociation and regeneration of quarkonium states. Semiclassical transport equations such as the Boltzmann and Langevin equations have been widely applied~\cite{Chen:2017duy,Du:2017qkv,Yao:2017fuc,Yao:2018zze,Yao:2018sgn,Yao:2020xzw,Zhao:2021voa}, which neglect important quantum effects such as the quantum coherence/decoherence of the quarkonium wavefunction. Recently, the open quantum system framework has been used to describe the dynamics of quarkonium inside the QGP and deepened our understanding~\cite{Akamatsu:2014qsa,Katz:2015qja,Brambilla:2016wgg,Brambilla:2017zei,Kajimoto:2017rel,Blaizot:2017ypk,Blaizot:2018oev,Yao:2018nmy,Miura:2019ssi,Yao:2020eqy,Brambilla:2020qwo,Brambilla:2021wkt}. See Refs.~\cite{Akamatsu:2020ypb,Yao:2021lus} for recent reviews. In this conference proceeding, I will review recent progress in applying the open quantum system framework to study quarkonium dynamics inside the QGP. In Section~\ref{sect:open}, the open quantum system framework will be reviewed, with an emphasis on its application to quarkonium in the QGP. The Markovian time evolution equations, i.e., the Lindblad equations and their validity conditions will also be discussed. Then in Section~\ref{sect:correlator}, I will discuss certain chromoelectric field correlators of the QGP that encode the essential properties of the medium in the Lindblad equations for quarkonium. I will compare the chromoelectric field correlators for quarkonium with those for unbound heavy quarks from both the mathematical and physical aspects. Finally, I will draw the conclusions in Section~\ref{sect:conclusions}.

\section{Open Quantum Systems and Lindblad Equations}
\label{sect:open}
We assume the system of a heavy quark pair (subsystem) and the QGP (environment) at thermal equilibrium is described by the Hamiltonians
\be
H = H_{Q\bar{Q}} + H_{\rm QGP} + H_I \,,
\ee
where $H_I$ describes the interaction between the heavy quark pair and the QGP. The density matrix of the whole system evolves in time according to the von Neumann equation
\be
\frac{\diff \rho(t)}{\diff t} = -i[H, \rho(t)] \,,
\ee
whose solution in the interaction picture can be symbolically written as
\be
\rho^{(\text{int})}(t) &=& U(t,0) \rho^{(\text{int})}(0) U^\dagger(t,0) \\
U(t) &=& \ml{T} \exp\Big(-i \int_0^t H_I^{(\text{int})}(t') \diff t'\Big) \,,
\ee
where $\ml{T}$ denotes the time-ordering operator. The time evolution of the subsystem consisting of the heavy quark pair can be written as
\be
\label{eqn:rho_s}
\rho_{Q\bar{Q}}^{(\text{int})}(t) &=& {\rm Tr}_{\rm QGP} \big( U(t,0) \rho^{(\text{int})}(0) U^\dagger(t,0) \big) \,.
\ee

The time evolution of the subsystem governed by Eq.~(\ref{eqn:rho_s}) is complicated in general. It can be greatly simplified in two limits: the quantum Brownian motion limit and the quantum optical limit. The two limits are valid under different hierarchies of time scales. Three time scales are relevant here: the environment (QGP) correlation time $\tau_E$ that determines the time domain of correlation functions of the environment, the subsystem (the heavy quark pair) intrinsic time scale $\tau_S$ and the subsystem relaxation time $\tau_R$. For the QGP at thermal equilibrium, it is expected $\tau_E\sim1/T$. The subsystem intrinsic time $\tau_S$ is estimated by the inverse of the typical energy gap in the subsystem and the relaxation time $\tau_R$ depends on the interaction strength between the subsystem and the environment. The quantum Brownian motion limit is valid when $\tau_R \gg \tau_E$ and $\tau_S \gg \tau_E$ while the quantum optical limit is valid when $\tau_R \gg \tau_E$ and $\tau_R \gg \tau_S$. The hierarchy $\tau_R \gg \tau_E$ is also called the Markovian limit in which there is no memory effect in the dynamics: the environment correlation has been lost during the subsystem relaxation, which is generally valid in the weak coupling limit, i.e., $H_I$ is weak.

Each limit corresponds to a different way of approximating the time integrals in Eq.~(\ref{eqn:rho_s}) under weak coupling expansion. In either limit, the time evolution of the subsystem density matrix can be written in terms of a Lindblad equation in the general form
\be
\frac{\diff \rho_{Q\bar{Q}}^{(\text{int})}(t)}{\diff t} = -i\big[ H_{Q\bar{Q},{\rm eff}}, \rho_{Q\bar{Q}}^{(\text{int})}(t) \big] + \sum_n D(n) \Big( L_n \rho_{Q\bar{Q}}^{(\text{int})}(t) L_n^\dagger - \frac{1}{2} \big\{ L_n^\dagger L_n, \rho_{Q\bar{Q}}^{(\text{int})}(t) \big\} \Big) \,,
\ee
where $L_n$'s are Lindblad operators whose explicit expressions depend on the effective field theory used to describe the heavy quark pair. Here $n$ denotes the quantum numbers of the heavy quark pair including their spatial positions. The term $D(n)$ can in principle be absorbed into the definition of the Lindblad operators. It is kept explicit here to emphasize that it is the only piece in the non-Hermitian parts that is related to the correlation functions of the environment, which will be discussed in more detail in the next section.

Details of the Lindblad equations in both limits and the justification of the hierarchies of the time scales for quarkonium in the QGP can be found in the recent review~\cite{Yao:2021lus}, as well as the connection between the Lindblad equations and the semiclassical transport equations.

\section{Chromoelectric field correlators of QGP}
\label{sect:correlator}
In either the quantum Brownian motion or the quantum optical limit, the QGP properties relevant to the small-size quarkonium in-medium dynamics are encoded in terms of correlators of chromoelectric fields, dressed with Wilson lines. The most general expressions of these gauge-invariant chromoelectric field correlators can be written as~\cite{Yao:2020eqy}
\begin{align}
[g_E^{++}]^{>}(q) &= \int\diff^4(y-x)\, e^{iq\cdot(y-x)} \Big\langle  \big[{E}_i(y) \ml{W}_{[( y^0, {\bs y}), (+\infty, {\bs y})]} \ml{W}_{[(+\infty, {\bs y}), (+\infty, {\bs \infty})]} \big]^a \nn\\
&\quad\quad\quad\quad\quad\quad\quad\quad\quad\quad \times \big[\ml{W}_{[(+\infty, {\bs \infty}), (+\infty, {\bs x})]}
 \ml{W}_{[(+\infty, {\bs x}),(x^0, {\bs x})]} {E}_i(x) \big]^a \Big\rangle_T \\
[g_E^{--}]^{>}(q) &= \int\diff^4(y-x)\, e^{iq\cdot(y-x)} \Big\langle   \big[\ml{W}_{[(-\infty, {\bs \infty}), (-\infty, {\bs y})]}
 \ml{W}_{[(-\infty, {\bs y}),(y^0, {\bs y})]} {E}_i(y) \big]^a \nn\\
&\quad\quad\quad\quad\quad\quad\quad\quad\quad\quad \times \big[{E}_i(x) \ml{W}_{[( x^0, {\bs x}), (-\infty, {\bs x})]} \ml{W}_{[(-\infty, {\bs x}), (-\infty, {\bs \infty})]} \big]^a \Big\rangle_T \,,
\end{align}
in which the chromoelectric fields are connected via staple-shape Wilson lines in adjoint representation. In $[g_E^{++}]^{>}$, the time-like Wilson lines extend to positive infinite time, accounting for the final-state interaction in quarkonium dissociation (the coupling between the center-of-mass motion of the octet $Q\bar{Q}$ pair and the QGP) while in $[g_E^{--}]^{>}$, the Wilson lines come from negative infinite time, representing the initial-state interaction in recombination. For the $Q\bar{Q}$ to thermalize properly, it is necessary for these two correlators to satisfy the Kubo-Martin-Schwinger (KMS) relation. The proof of the KMS relation is more involved here than textbook examples because of the Wilson lines. The KMS relation was first proved in Ref.~\cite{Binder:2021otw} by using the parity and time-reversal transformations and assuming the density matrix of the QGP at thermal equilibrium is invariant under these transformations.

These chromoelectric field correlators are momentum dependent and appear in differential reaction processes. For inclusive reaction processes, one integrates over the momentum exchanged and obtain the momentum independent correlators
\be
\label{eqn:G}
G^>_E(q_0) = \int \frac{\diff^3{\bs q}}{(2\pi)^3}
[g_E^{++}]^{>}(q) = \int \frac{\diff^3{\bs q}}{(2\pi)^3} [g_E^{--}]^{>}(q) = \int \diff t\, e^{iq_0t} 
\left\langle {E}^a_i(t) \ml{W}^{ab}(t,0)  {E}^b_i(0) \right\rangle_T \,, \ \ \
\ee
where we have used translational invariance in time. In the quantum Brownian motion limit, the hierarchy of time scales $\tau_S \gg \tau_E$ leads to $T \gg H_{Q\bar{Q}}$ which justifies an expansion in terms of $H_{Q\bar{Q}}/T$. Physically this means the binding energy effect is negligible in quarkonium dynamics. Therefore in the quantum Brownian motion limit, only the zero frequency limit of the chromoelectric field correlator contributes, i.e., $G^>_E(q_0=0)$. On the other hand, no hierarchy exists between the QGP temperature and the binding energy in the quantum optical limit. As a result, binding energy effect cannot be neglected, which can suppress the reaction rates~\cite{Blaizot:2021xqa} and is important for the quarkonium states to thermalize properly~\cite{Yao:2017fuc}. Due to energy conservation in this limit, only the finite frequency part of the chromoelectric field correlator matters in bound-unbound transitions.

Finally, it is worth noting that the correlator (\ref{eqn:G}) is similar to but different from the correlator used to define the heavy quark diffusion coefficient~\cite{Casalderrey-Solana:2006fio}
\be
\label{eqn:kappa}
\kappa = \lim_{q_0\to0} \int \diff t\, e^{iq_0t}
\Big\langle {\rm Tr}_{\rm color} \big[ U(-\infty,t) E_i(t) U(t,0) E_i(0) U(0,-\infty) \big] \Big\rangle_T \,,
\ee
where the Wilson line is in fundamental representation. The two correlators (\ref{eqn:G}) and (\ref{eqn:kappa}) differ in the Wilson line configurations. In the dynamics of open heavy quarks, the heavy quark carries color all the way, so the Wilson lines extend to both positive and negative infinite times. Perturbative calculations~\cite{Burnier:2010rp,Binder:2021otw} showed that these two correlators agree in the temperature dependent part but differ by a finite constant at next-to-leading order. See also the discussions in Ref.~\cite{Eller:2019spw}.

\section{Conclusions}
\label{sect:conclusions}
I reviewed recent progress in applying the open quantum system framework to study quarkonium dynamics inside the QGP, with a focus on the QGP correlators of the chromoelectric fields that enter the Lindblad equations in both the quantum Brownian motion and the quantum optical limits. The same framework has also been applied to study Dark Matter bound state formation in the coannihilation scenario in the early universe~\cite{Binder:2021otw}. Some problems of jet quenching in heavy ion collisions have also been studied in the open quantum system framework~\cite{Vaidya:2020cyi,Vaidya:2020lih,Vaidya:2021vxu}. Due to the importance of the Lindblad equation in heavy ion physics and the high computational cost to solve it, it is worth exploring to solve it on quantum computers. This has been explored for a toy model in the quantum optical limit~\cite{DeJong:2020riy} and for the U(1) gauge theory in 1+1 dimension (also known as the Schwinger model) in the quantum Brownian motion limit~\cite{deJong:2021wsd}. It is expected that advanced quantum algorithms and machine learning techniques may significantly speed up simulating open quantum systems in the near future and deepen our understanding of physics in heavy ion collisions.

\acknowledgements
This work is supported by the U.S. Department of Energy, Office of Science, Office of Nuclear Physics grant DE-SC0011090.

\bibliographystyle{apsrev4-1}

\end{document}